\documentclass[aip,amsmath,amssymb,reprint,epsfig]{revtex4-1}

\usepackage{graphicx}
\usepackage{dcolumn}
\usepackage{bm}
\usepackage[utf8]{inputenc}
\usepackage[T1]{fontenc}
\usepackage{mathptmx}
\usepackage{etoolbox}
\usepackage{amsmath}
\usepackage{amssymb}
\usepackage{color}

\makeatletter
\def\@email#1#2{%
 \endgroup
 \patchcmd{\titleblock@produce}
  {\frontmatter@RRAPformat}
  {\frontmatter@RRAPformat{\produce@RRAP{*#1\href{mailto:#2}{#2}}}\frontmatter@RRAPformat}
  {}{}
}%
\makeatother
\begin{document}

\preprint{AIP/123-QED}

\title[Convective Granular Flows]{Convective Flows in Sheared Packings of Spherical Particles}
\author{Mehran Erfanifam}
\affiliation{Department of Physics, Institute for Advanced Studies in Basic Sciences (IASBS), 
Zanjan 45137-66731, Iran.}
\author{Mahnoush Madani}
\affiliation{Faculty of Mathematics and Natural Sciences, Heinrich Heine Universit\"at 
D\"usseldorf, 40225 D\"usseldorf, Germany.}
\author{Reza Shaebani}
\email{shaebani@lusi.uni-sb.de.}
\affiliation{Department of Theoretical Physics and Center for Biophysics, Saarland University, 
66123 Saarbr\"ucken, Germany.}
\author{Maniya Maleki}
\email{maniya.maleki@uni.edu.}
\affiliation{Department of Physics, Institute for Advanced Studies in Basic 
Sciences (IASBS), Zanjan 45137-66731, Iran.}
\affiliation{Department of Chemistry, University of Pennsylvania, Philadelphia, Pennsylvania 
19104, USA.}
\affiliation{Department of Physics, University of Northern Iowa, Cedar Falls, IA 50614, USA.}

\date{\today}
  
\begin{abstract}
Understanding how granular materials respond to shear stress remains a central 
challenge in soft matter physics. We report direct observations of persistent 
granular convection in the bulk shear zones of spherical particle packings--- 
a phenomenon previously associated primarily with particle shape anisotropy 
or boundary effects. By employing various bead-coloring techniques in a 
split-bottom geometry, we reveal internal flow fields within sheared granular 
packings. We find robust convection rolls, strikingly governed by system 
geometry: at low filling heights, two counter-rotating convection rolls emerge, 
while at higher filling heights, a single dominant convection roll forms, 
featuring radially outward flow at the surface. This transition is driven by 
the height-dependent broadening of the shear zone, which introduces shear rate 
asymmetry across its flanks. Notably, the transition occurs entirely within 
the open shear band regime. These findings underscore the pivotal role of 
system geometry in shaping secondary flow formation in dense packings of 
frictional particles, suggesting possible broader relevance to geophysical 
flow dynamics and industrial applications.
\end{abstract}

\maketitle

Granular flow under shear presents fundamental challenges with wide-ranging 
implications, from powder processing to geological faulting. In slowly sheared 
granular systems, shear strain localizes into shear zones--- often narrow regions 
near boundaries, but also broad zones within the bulk \cite{Schall10,Mueth00,
Kollmer20,Fenistein03,Shaebani21,Cheng06,Unger04,Torok07,Moosavi13,Henann13,
Artoni18}. While such primary, shear-aligned flows have been extensively 
characterized, the emergence of secondary flows--- circulatory motions transverse 
to the main shear direction--- remains less understood. Although typically weaker 
than the primary flow, secondary convective flows can strongly influence bulk 
transport, particle mixing, and segregation \cite{Harrington13,Fan10,khosropour97}, 
making them crucial in both natural and industrial contexts.

Convection rolls in granular flows are often attributed to narrow shear zones 
near boundaries, such as in Couette geometries, where the interplay of shear-induced 
dilatancy, gravity, and centrifugal forces drives circulatory motion in radial 
and vertical directions \cite{Murdoch13,Krishnaraj16,Zheng20,khosropour97}. 
However, many natural and industrial flows involve spatially extended shear 
zones that are not confined to boundaries. This raises a key question:\ Is 
the generation of secondary flows a generic feature of sheared granular 
matter, or does it fundamentally require boundaries? Split-bottom 
geometries--- designed to probe wide shear zones away from lateral walls 
\cite{Fenistein03,Shaebani21,Cheng06,Unger04,Torok07,Moosavi13,Henann13,
Harrington13}--- provide a powerful platform to address this. In such 
systems, convective flows have been observed in binary mixtures 
\cite{Harrington13,Fan10}, and surface heaping accompanied by secondary 
flows was reported for highly elongated particles \cite{Fischer16,Wortel15}, 
where misalignment between particle orientation and flow streamlines was 
proposed as the driving mechanism. In contrast, spherical particles exhibit 
no surface heaping, reinforcing the prevailing assumption that significant 
secondary convection is absent. While recent simulations suggest possible 
secondary motion in split-bottom geometries \cite{Dsouza21}, direct experimental 
evidence for persistent secondary flows in spherical particle packings far from 
boundaries has remained elusive.

\begin{figure}[b]
\centering
\includegraphics[width=0.47\textwidth]{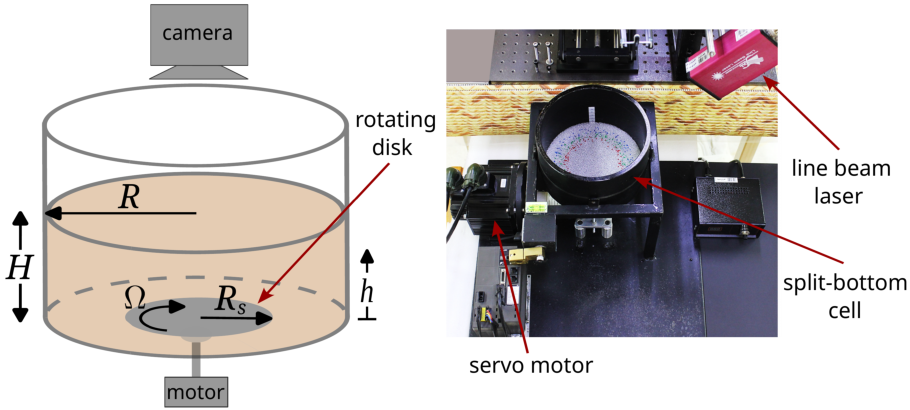}
\caption{Schematic (left) and photograph (right) of the split-bottom setup 
used in the experiments.}
\label{Fig1}
\end{figure}

\begin{figure*}[t]
\centering
\includegraphics[width=0.99\textwidth]{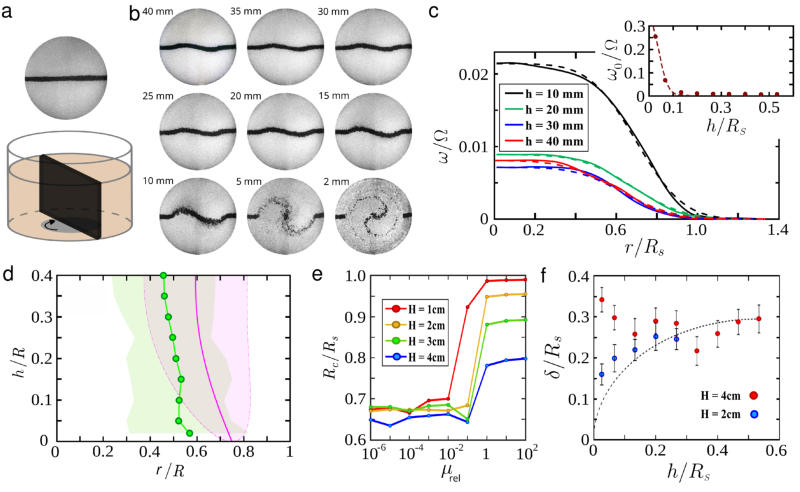}
\caption{(a) Schematic of initial vertical plane of colored tracers embedded along one 
diameter of the cylinder and initial surface photo before shearing. (b) Deformed tracer 
line after shear reveals local angular displacement at different bulk heights $h$. (c) 
Angular velocity profiles $\omega(r)$ at various $h$ (solid lines) and corresponding 
error function fits (dashed lines). Inset: axial angular velocity $\omega_0$ vs $h$. 
The line represents a Gaussian fit. (d) Comparison of shear zone center $R_c$ and width 
$\delta$ in our setup (green) with prior rough-boundary data (purple). Solid lines and 
symbols represent $R_c$ and shaded regions indicate $\delta$; see text. (e) Variational 
model prediction of $R_c$ versus $\mu_\text{rel}$ for different $H$. (f) $\delta$ vs 
$h$ for different values of $H$. The dotted line represents $\delta{\sim}h^{0.58}$.}
\label{Fig2}
\end{figure*}

Here, we experimentally reveal that dense packings of spherical particles 
under shear can exhibit sustained bulk convection. Using a split-bottom 
cell combined with bead-coloring techniques, we visualize internal flow 
fields and uncover robust convection rolls deep within the bulk--- 
despite the absence of significant surface heaping, consistent with previous 
surface-based observations \cite{Fischer16,Sakaie08,Cabrera20,Mohammadi22}.  
Strikingly, we observe a geometry-controlled transition in convection 
patterns from counter-rotating rolls at low filling heights to a single 
dominant convection cell at higher heights. We show that this transition 
is driven by the filling-height-dependent broadening of the shear zone, 
which induces asymmetry in local shear rates across the flow. These 
findings demonstrate that even in the absence of particle anisotropy 
or boundaries, geometry alone can generate persistent secondary flows 
in spherical granular systems, offering fresh insights into shear-induced 
transport in dense granular media.

\begin{figure*}[t]
\centering
\includegraphics[width=0.99\textwidth]{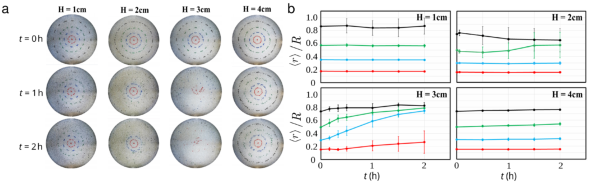}
\caption{(a) Top-view images at selected time points for various filling heights 
$H$, showing the dispersion of colored tracer grains initially placed in narrow 
radial bands on the surface. (b) Mean radial position $\langle r\rangle$ of 
tracers as a function of time, for different initial radii and filling heights. 
Changes in $\langle r\rangle$ reflect surface signatures of underlying convective flow.}
\label{Fig3}
\end{figure*}

\emph{Setup and procedure.---}
Our experiments use a split-bottom geometry consisting of a rotating disk of 
radius $R_s{=}7.5\,\text{cm}$ flush with a fixed outer ring inside a cylindrical 
cell of radius $R{=}10\,\text{cm}$ (Fig.\,\ref{Fig1}). The disk is rotated at a 
constant angular velocity $\Omega{=}0.127\,\text{rad/s}$ (${\simeq}\,1.213\,\text{rpm}$). 
The angular velocity of the disk is chosen sufficiently low to ensure operation 
in the slow quasistatic granular‐flow regime, in which the flow patterns are 
known to be essentially shear‐rate independent \cite{GDRMiDi04,Fenistein04}. The 
granular medium comprises $2\,\text{mm}$ glass beads with 15\% polydispersity, which 
has negligible effect on packing fabric \cite{Shaebani12b}. Colored tracer grains 
are used to visualize internal flow. All experiments employing colored tracers 
use the same base glass beads. The applied color coating adds only a few 
micrometers to the bead diameter, which is negligible compared with their size. 
The cylinder is filled to height $H$ with a bulk packing fraction of ${\sim}
\,60\%$. A standardized pouring protocol, with gently leveling the surface 
every ${\sim}\,8\,\text{cm}$, ensures reproducibility. Smooth chamber walls 
allow axial slip, promoting shear localization within the bulk. To reconstruct 
internal flow patterns, we photograph horizontal cross-sections from above 
after carefully removing successive top layers, yielding 2D tomographic 
maps of tracer positions within the bulk. We note that, despite the relatively 
large bead size, random packing readily produces smooth shear zones and 
desired filling heights with less than $1\%$ variation across the large 
cylinder. The presence of many beads in the shear zone and surface layer 
smooths surface fluctuations and minimizes finite-size effects. Using 
significantly smaller beads is not practical, as electrostatic charging 
and cohesive forces become relevant, causing the material to deviate 
from dry, cohesionless granular behavior.

\emph{Bulk primary flow profiles.---} 
To characterize the primary flow field, we first determine the bulk shear profile. 
To this aim, we design the initial conditions such that a vertical plane of 
black-colored tracer particles is embedded along one diameter of the cylinder, 
forming a rectangular sheet from the bottom to the free surface and spanning 
the full cylinder width [Fig.\,\ref{Fig2}(a)]. After shearing the system for 
$210\,\text{s}$, we photograph the surface, then sequentially remove ${\sim}\,
2\,\text{mm}$ slices from the top and image each exposed layer. This procedure 
yields the deformed shape of the initially straight tracer line at each bulk 
height $h$ [Fig.\,\ref{Fig2}(b)], from which local angular displacements and 
velocity profiles $\omega(r)$ at various depths can be reconstructed; see 
Fig.\,\ref{Fig2}(c) and Refs.\,\cite{Fenistein03,Shaebani21}. These profiles are 
well described by an error function $\omega{=}\frac{\omega_0}{2}\left[1{-}
\text{erf}\left(\frac{r{-}R_c}{\delta}\right)\right]$, allowing extraction 
of the axial angular velocity $\omega_0$ (inset of panel c) and the center 
position $R_c$ and width $\delta$ of the shear zone (panel d). Notably, 
$\omega_0$ is small even near the rotating base due to slip at the bottom 
boundary. Moreover, the bulk angular velocity decays rapidly with $h$. The 
quantitative difference between the profile at $h\,{=}\,10\,\text{mm}$ and 
those at larger heights is evident in Fig.\,\ref{Fig2}(c). The reduction 
of angular velocity with increasing $h$ is distinctly nonuniform. Starting 
from $h\,{=}\,0$, the shear zone shifts quickly toward the cylinder axis, 
accompanied by a rapid decrease in rotation speed. For heights above $h\,
{=}\,20\,\text{mm}$, however, the rate of change slows markedly. As shown 
in the inset of Fig.\,\ref{Fig2}(c), the rapid decay of axial angular 
velocity with $h$ follows a Gaussian trend \cite{Cheng06,Madani21}.

\begin{figure}
\centering
\includegraphics[width=0.47\textwidth]{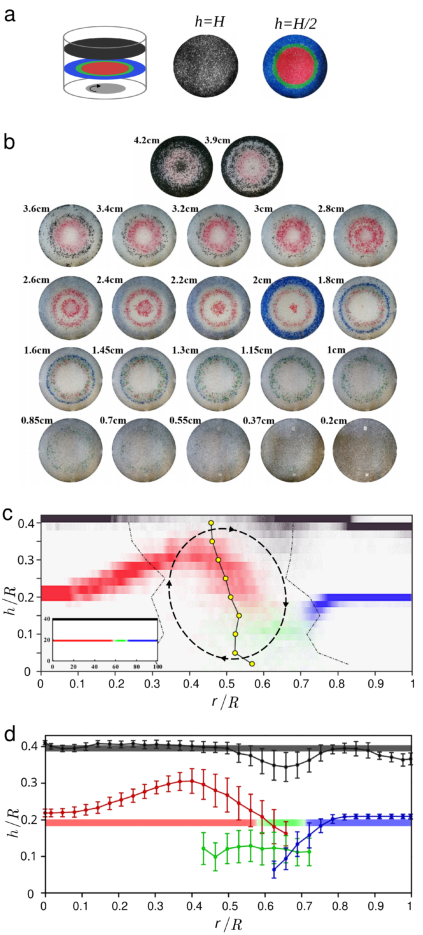}
\caption{(a) Initial configuration of colored tracers placed at mid-height ($h{=}H{/}2$) 
for two filling levels, covered with a black tracer layer at the top. (b) Reconstructed 
bulk tracer positions after shear for $H{=}4\,\text{cm}$. (c) Tracer density map in the 
($r, h$) plane for $H{=}4\,\text{cm}$ in the steady state, overlaid with the shear zone 
center and boundaries, indicating a single dominant convection roll. Inset: Side view 
of the initial arrangement of colored tracers. (d) Quantitative characterization of 
the vertical positions of the differently colored tracer particles shown in panel (c). 
Each point denotes the mean color-intensity-weighted position within a vertical 
strip and the error bars indicate the corresponding standard deviations.}
\label{Fig4}
\end{figure}

\begin{figure}
\centering
\includegraphics[width=0.4\textwidth]{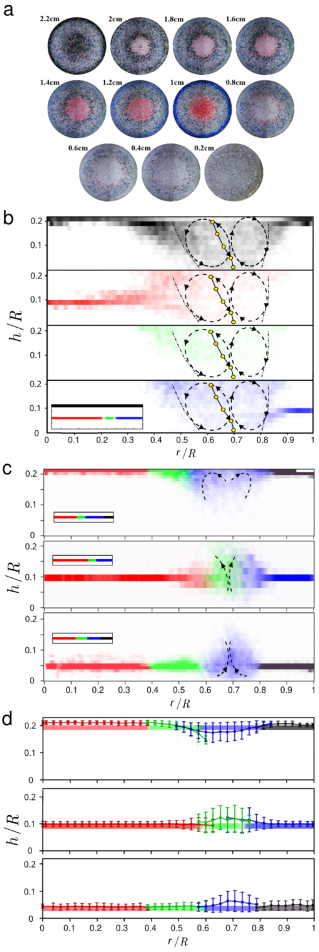}
\caption{(a) Reconstructed bulk tracer positions after shear for $H{=}2\,\text{cm}$. (b) 
Individual tracer density maps for $H{=}2\,\text{cm}$ in the steady state, overlaid with 
the shear zone center and boundaries, indicating two counter-rotating convection rolls. 
Inset: Side view of the initial arrangement of colored tracers. (c) Tracer density maps 
for $H{=}2\,\text{cm}$ and bulk heights $h{=}0.5,\,1.0,\,2.0\,\text{cm}$ at an early 
time point $t\,{=}\,3\,\text{min}$. (d) Mean vertical positions 
(color-intensity-weighted) of the differently colored tracer particles shown in panel (c).}
\label{Fig5}
\end{figure}

Comparing $R_c$ and $\delta$ in our setup with known results in similar but rough-bottom 
geometries--- where the behavior is well captured by $h{=}H{-}R_c\big[1{-}\frac{R_s}{R_c}
\big(1{-}(\frac{H}{R_s})^{5/2}\big)\big]^{2/5}$ and $\delta{\sim}h^{\beta}$ ($0.5{<}\beta
{<}1$) \cite{Fenistein03,Unger04,Torok07}--- reveals significant deviations in $R_c$ and 
$\delta$ due to slip [Fig.\,\ref{Fig2}(d)]. We confirm this through numerical minimization 
of energy dissipation using a variational approach \cite{Unger04,Torok07,Shaebani21,Madani21,
Unger07}, showing that reduced bottom friction--- characterized by an effective friction 
$\mu_\text{rel}{=}\mu_\text{bottom}^\text{eff}{/}\mu_\text{bulk}^\text{eff}$--- shifts 
$R_c$ inward [Fig.\,\ref{Fig2}(e)]. As shown in Fig.\,\ref{Fig2}(f), shear zone broadening 
due to slip enhances with filling height--- which increases shear rate asymmetry across 
the flanks, an essential ingredient for the emergence of convection. These results justify 
our choice of a smooth-bottom setup to access broader shear zones and thereby amplify the 
secondary flow signatures.

\emph{Surface radial flow.---} 
To probe secondary flow near the surface, we place colored tracer grains (red, blue, green, 
black) in narrow radial bands at different radii atop the transparent bed [Fig.\,\ref{Fig3}(a)] 
and track their radial displacement over time. The mean tracer radius $\langle r\rangle$ for 
each color is monitored at 30-minute intervals for various filling heights $H{=}1{-}4\,\text{cm}$. 
At $H{=}1\,\text{cm}$, only black tracers near the outer edge ($r{=}85\,\text{mm}$) exhibit 
slight spreading and other colors remain stationary, indicating negligible surface convection 
[Fig.\,\ref{Fig3}(b)]. 

At $H{=}2\,\text{cm}$, tracers at radii $r{=}50,\,75\,\text{mm}$ (near 
the flanks of the surface shear zone) disperse and converge toward an intermediate radius, 
while inner tracers at $r{=}15,\,30\,\text{mm}$ remain fixed. 
This convergence indicates the presence of two counter-rotating convection rolls, featuring either downward flow along the shear flanks and upward return flow between them, or the reverse pattern. During the initial stage (up to $t = 3~\mathrm{min}$), black particles at $r = 75~\mathrm{mm}$ move outward, while green particles at $r = 50~\mathrm{mm}$ move inward. Subsequently, the particles in this region descend through the convection rolls and re-emerge at the surface at radii different from their initial positions, leading to a shift in their mean radii in the opposite direction [Fig.\,\ref{Fig3}(b)]. The initial divergence and subsequent convergence of the black and green particles indicate a downward flow along the shear flanks and an upward return flow between them.

For $H{=}3\,\text{cm}$, the surface flow is strongest: tracers at $r{=}30,\,50\,\text{mm}$ move outward up to radial 
distance $r{\approx}70\,\text{mm}$, consistent with a dominant single convection roll with 
downward flow at the outer shear flank at $r{\simeq}70\,\text{mm}$ and upward return flow 
at the inner shear flank at $r{\simeq}30\,\text{mm}$. Tracers at $r{=}15,\,75\,\text{mm}$ 
show minimal displacement, indicating they lie outside the active convective region.

 At $H{=}4\,\text{cm}$, surface flow is largely suppressed, though there still exists a weak 
radial motion. The peak radial surface velocity $v_r\,{\approx}\,6{\times}10^{-4}\,\text{mm/s}$ 
is about 2\% of the peak azimuthal speed $v_\theta\,{\approx}\,3{\times}10^{-2}\,\text{mm/s}$, 
reflecting the dynamical relevance of convective flows at this height.

\emph{Bulk convective flow.---} 
To probe internal convection, we prepare a single colored layer at mid-depth ($h{=}H{/}2$) for 
two filling heights $H{=}2,\,4\,\text{cm}$. Tracers are arranged in three radial bands: red 
beads in the inner region ($r{<}6\,\text{cm}$), green in a ring ($6{<}r{<}7\,\text{cm}$), and 
blue in the outer bulk ($7{<}r{<}10\,\text{cm}$), with a top layer of black tracers covering 
the free surface [Fig.\,\ref{Fig4}(a)]. After shearing the system for 60 min ($H{=}2\,\text{cm}$) 
or 180 min ($H{=}4\,\text{cm}$), the system is excavated layer by layer, and the bulk tracer 
positions are reconstructed [Figs.\,\ref{Fig4}(b) and \ref{Fig5}(a)]. For $H{=}4\,\text{cm}$, 
the tracer density map in the ($r, h$) plane shown in Fig.\,\ref{Fig4}(c,d) reveals a strong single 
convection roll:\ red tracers rise while green and blue tracers descend within the shear zone. 
Black surface tracers are displaced outward and drawn into the bulk at the outer flank. A small 
surface elevation in the central region suggests dilatancy induced by flow. The entire convective 
motion is confined within the shear zone, with negligible transport beyond its boundaries. 

For $H{=}2\,\text{cm}$, tracer trajectories [Fig.\,\ref{Fig5}(b)] reveal that the flow direction 
of the two counter-rotating convection rolls is downward at the shear flanks: Tracers from all 
colors descend along bulk shear flanks, reappear at the top free surface, and move toward the 
edges of the shear zone, consistent with the surface flow observations. Black surface tracers 
are drawn toward the surface shear zone edges and disappear into the bulk, further corroborating 
the counter-rotating rolls. Formation of two counter-rotating convection rolls for $H{=}2\,\text{cm}$ 
is also evident when we track the colored beads at an earlier time point $t\,{=}\,3\,\text{min}$ 
for bulk heights $h{=}0.5,\,1.0,\,2.0\,\text{cm}$; see Fig.\,\ref{Fig5}(c,d). Overall, convective 
transport is localized within the shear zone and highly sensitive to the filling height.

\begin{figure}[t]
\centering
\includegraphics[width=0.47\textwidth]{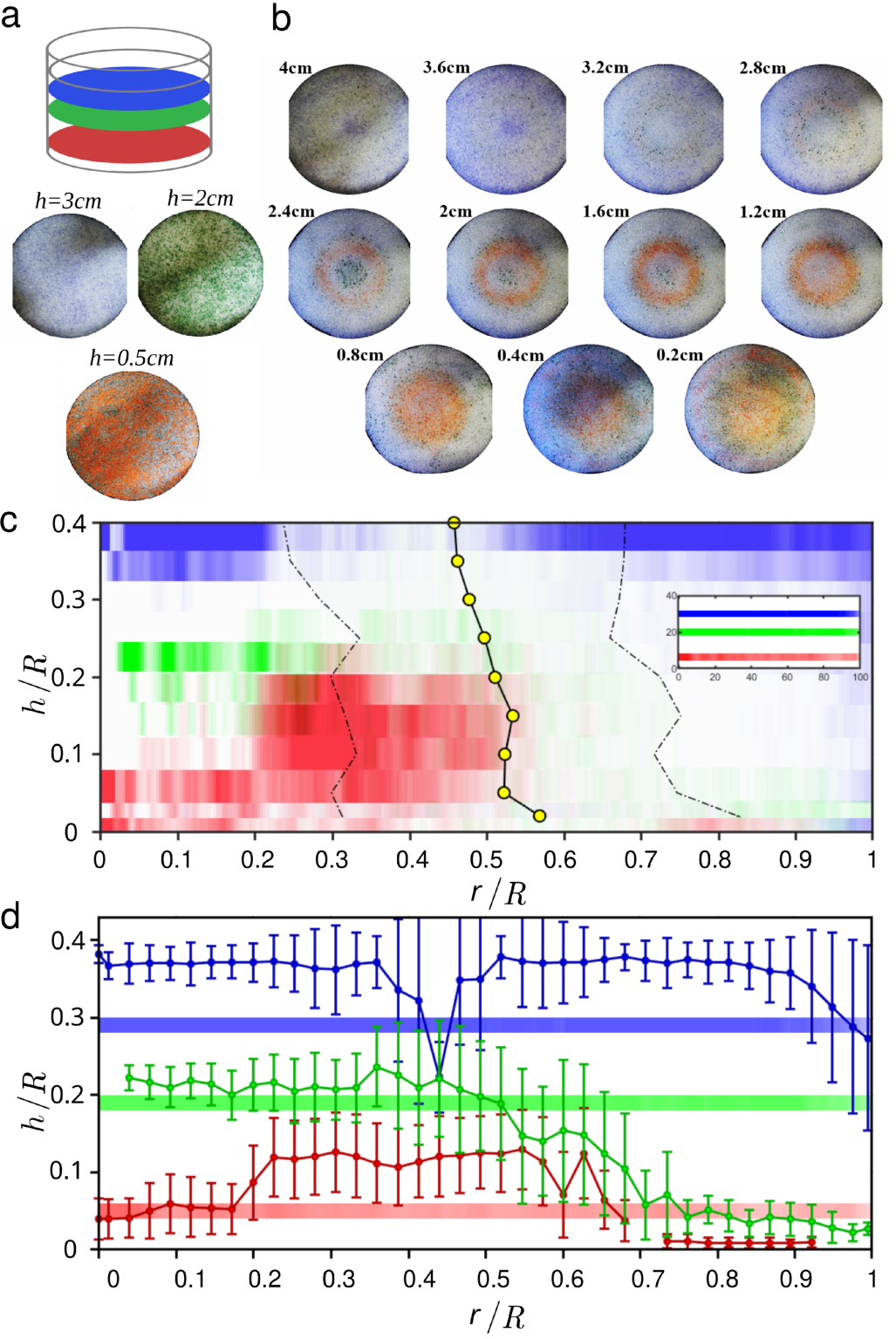}
\caption{(a) Initial configuration of horizontal red, green, and blue tracer layers at depths 
$h{=}0.5,\,2,\,3\,\text{cm}$, respectively, in a granular pile of height $H{=}4\,\text{cm}$. 
(b) Reconstructed positions of the tracers after 160 minutes of shear. (c) 2D tracer density 
map in the ($r, h$) plane, showing radial and vertical displacements. Shear band center and 
boundaries are overlaid. Inset: Side view of the initial arrangement of colored tracers. (d) 
Mean vertical positions (color-intensity-weighted) of the differently colored tracer particles 
shown in panel (c).}
\label{Fig6}
\end{figure}

To further improve the bulk resolution at large filling height $H{=}4\,\text{cm}$, we implement 
a layered tracer configuration. Colored grains are arranged in horizontal planes:\ red at $h{=}
0.5\,\text{cm}$, green at $h{=}2\,\text{cm}$, and blue at $h{=}3\,\text{cm}$ [Fig.\,\ref{Fig6}(a)]. 
After shearing the system for 160 mins, we excavate the packing layer by layer, capturing the 
tracer positions at each step [Fig.\,\ref{Fig6}(b)]. This procedure enables us to reconstruct 
a 2D density map of tracer positions in the ($r, h$) plane [Fig.\,\ref{Fig6}(c,d)]. The tracer 
displacements reveal clear radial and vertical motions indicative of secondary flow in the bulk. 
Red particles near the base migrate inward and then upward, tracing a clockwise circulation. 
Green tracers from the mid-plane predominantly descend and accumulate near the bottom within 
the shear zone, while a portion at smaller radii is lifted upward--- again consistent with 
toroidal motion. Blue tracers, initially near the top, move downward at larger radii. The overall 
pattern confirms the presence of a dominant single convection roll confined mainly within 
the bulk shear zone.

\begin{figure*}[t]
\centering
\includegraphics[width=0.85\textwidth]{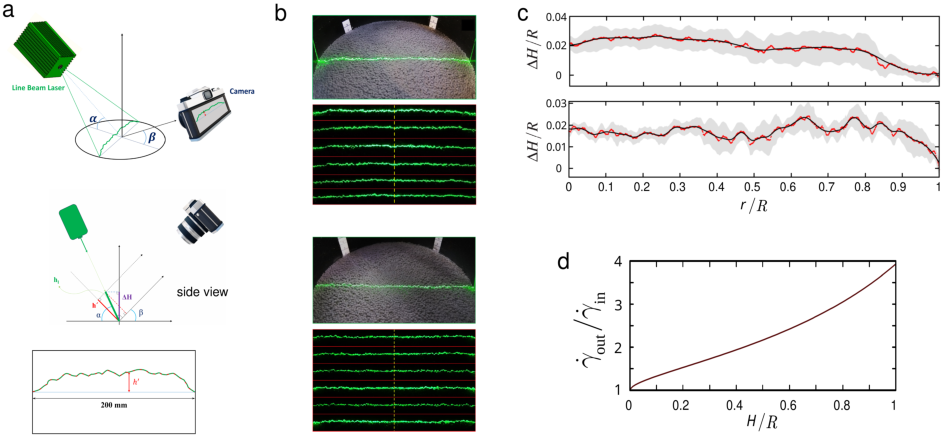}
\caption{(a) Schematic of the laser-sheet setup showing the illumination geometry and camera 
orientation, together with a side-view illustration of the oblique laser projection on the 
granular surface. (b) Example images of the ash-coated free surface illuminated by the laser 
line, along with six extracted profiles corresponding to the six measured diameters for the 
filling height $H\,{=}\,4\,\text{cm}$ (top) and $H\,{=}\,2\,\text{cm}$ (bottom). (c) Azimuthally 
averaged surface height-change profiles (relative to the initial flat surface) for $H\,{=}\,
4\,\text{cm}$ (top) and $H\,{=}\,2\,\text{cm}$ (bottom). Raw digitized data (red points) and 
smoothed curves (black lines) are shown. (d) Fraction of shear rates at the outer and inner 
flanks of the surface shear zone vs $H$. The asymmetry grows with filling height, promoting 
a transition from symmetric counter-rotating convection rolls to a dominant single roll 
as shear becomes localized more strongly at the outer edge.}
\label{Fig7}
\end{figure*}

\emph{Dilatancy and heap formation.---}
To quantify surface deformation and dilatancy during shear, we employ a 532.8 nm laser line 
(Fath Optics) projected at an oblique angle $\alpha\,{=}\,65^\circ$ relative to the horizontal,
illuminating a diagonal path across the granular free surface [Fig.\,\ref{Fig7}(a)]. Because 
the transparent glass beads scatter light and broaden the laser sheet, a thin opaque ash layer 
(thickness ${<}\,75\,\mu\text{m}$) is applied to the surface to enhance optical contrast 
without measurably altering surface roughness. A camera positioned at a fixed angle $\beta
\,{=}\,45^\circ$ to the horizontal captures images of the illuminated line. For each filling 
height, six diameters are recorded by rotating the laser-camera assembly around the cylinder 
axis [see Fig.\,\ref{Fig7}(b)].

From each image, the apparent pixel height $h'$ of the illuminated line is extracted using 
custom MATLAB routines. The true vertical height deviation $\Delta H$ is then obtained 
through a geometric correction based on the known projection and viewing angles as 
$\Delta H\,{=}\,h'\,\frac{\sin\alpha}{\sin(\alpha{+}\beta)}$. The resulting six profiles 
are azimuthally averaged to obtain the mean surface-height change, with the standard 
deviation indicated as shaded envelopes in Fig.\,\ref{Fig7}(c).

The optical resolution of the method depends on the field of view and ranges between 
0.06-0.11 mm, sufficient to resolve millimeter-scale heap features given 
the ${\sim}1$ mm laser-line width. Calibration using reference glass beads of diameter 
2 mm yields an accuracy of ${\pm}\,0.05$ mm (${\sim}\,5\%$ relative error), consistent 
across filling heights. Tests on smooth reference surfaces with and without the ash 
layer show ${<}5\%$ variation in measured $\Delta H$, confirming negligible influence 
from ash-induced micro-roughness.

For clarity of presentation, the raw profiles (red points) are shown together with 
smoothed curves obtained via robust LOESS filtering (span 0.04 for $H\,{=}\,2\,\text{cm}$ 
and 0.08 for $H\,{=}\,4\,\text{cm}$), which suppress digitization noise and grain-scale 
curvature while preserving physically meaningful variations. As shown in Fig.\,\ref{Fig7}(c) 
for $H\,{=}\,2\,\text{cm}$, a pronounced elevation develops within the shear zone 
($r\,{\approx}\,50{-}75$ mm), reaching a maximum $\Delta H\,{\approx}\,2.2\,\text{mm}$ 
(${\sim}11\%$ of $H$). For $H\,{=}\,4\,\text{cm}$, the surface exhibits a broader central 
rise of $\Delta H\,{\approx}\,2.5\,\text{mm}$ (${\sim}7\%$ of $H$) within the shear zone. 
Although these dilatancy-induced height changes are modest compared with the large heaps 
commonly reported for elongated particles \cite{Fischer16,Wortel15}, their magnitude and 
radial position are fully consistent with the emergence of a single roll or a pair of 
counter-rotating convection rolls at different filling heights.

\emph{Onset of convection roll transition.---}
To rationalize the transition from two counter-rotating convection rolls at low filling heights 
to a single dominant roll at larger fillings, we examine how the radial shear-rate distribution 
evolves with the broadening of the shear zone. Using the error-function form of the surface 
angular-velocity profile $\omega(r)$, the associated tangential velocity is $v_\theta(r)
{=}\,r\,\omega(r)$, from which the radial shear rate is obtained as $\dot\gamma(r){=}
\frac12(\partial_r v_\theta{-}v_\theta{/}r)=\frac12 r \partial_r \omega$. The shear-rate 
profile exhibits two peaks located around the flanks of the shear zone (defined 
as $R_c{\pm}\delta$). We estimate the shear rates at these two locations and plot their 
ratio $\dot\gamma_\text{out}{/}\dot\gamma_\text{in}$ as a function of filling height 
$H{\propto}\,\delta^{1{/}\beta}$ (with $\beta{=}0.5$) in Fig.\,\ref{Fig7}(d). 
This ratio increases monotonically with $H$, indicating a growing shear-rate imbalance 
between the outer and inner flanks.

To connect this shear-rate asymmetry to the emergence of secondary convection, we assume a 
monotonic relation between the local shear rate and local dilation \cite{khosropour97}. Thus, 
large dilatancies occur around the two peaks of shear rate located at the flanks of the shear 
zone. At small filling heights, the two shear-rate peaks are nearly symmetric, and the 
corresponding dilatancy regions, where the material locally expands and becomes more mobile, 
are of comparable strength. Because gravity biases particle motion downward in locally dilated 
regions, both flanks support similarly strong downward flow components, resulting in the 
coexistence of two counter-rotating convection rolls. As the filling height increases, the 
shear zone widens and the shear-rate profile becomes increasingly asymmetric:\ the outer flank 
experiences significantly stronger shear, producing a stronger downward driving component for 
convection. Enhanced dilatancy reduces local packing fraction and facilitates particle 
rearrangements, allowing gravity-driven downward motion to dominate on the outer side. 
The inner flank, experiencing weaker shear and weaker dilatancy, can no longer sustain 
an opposing convective roll. This imbalance naturally yields the experimentally observed 
transition to a single, outer-driven roll at larger filling heights. Our measured surface-height 
profiles shown in Fig.\,\ref{Fig7}(c) are consistent with this model; the profiles reveal 
dilatancy-induced elevations whose magnitude and radial location correlate with the proposed 
single roll or two counter-rotating convection rolls for $H\,{=}\,4$ or $2\,\text{cm}$. 

This qualitative framework, linking the filling-height-dependent imbalance 
between the two shear-rate peaks with gravity-biased motion in locally dilated regions, 
provides a consistent minimal mechanism for both (i) the formation of secondary convection 
rolls in spherical granular packings and (ii) their transition from two rolls to a single 
roll as $H$ increases. It also suggests that similar transitions might occur in tall granular 
piles. At large depths, however, gravity-induced confining stresses suppress particle 
rearrangements \cite{Shaebani07,Shaebani08} and thereby inhibit the formation of convection 
rolls. Under reduced-gravity conditions \cite{Murdoch13,Born17}, this constraint is partially 
relaxed. Yet a lower gravitational acceleration also weakens the gravity-driven downward 
motion that is essential for sustaining convection. One may further argue that reduced 
gravity diminishes frictional interparticle forces, leading to reduced dilatancy and thus 
an additional weakening of secondary convection. Consistent with these expectations, 
experiments have indeed reported suppressed convective flows under microgravity \cite{Murdoch13}.

We note that granular flows in the quasistatic regime, as studied here, are fully reproducible. For 
a given filling height and shear protocol, both the primary flow field and the secondary convection 
pattern remain robust and do not vary between experimental runs. The error bars presented in our 
analysis therefore reflect intrinsic tracer dispersion, tracer diffusion, and finite particle-size 
effects, as the dominant sources of experimental variability in our system.

We have presented experimental evidence of geometry-driven convective flow in sheared 
granular packings of spherical particles. Using tracer layers and tomographic reconstruction, we 
reveal a transition from symmetric counter-rotating convection rolls at low filling heights to a 
dominant single-roll pattern at higher fillings. This transition correlates with increasing asymmetry 
in the radial shear profile, consistent with a shear-driven convection mechanism. Our work provides 
direct experimental evidence that secondary convection rolls can arise in the bulk of spherical 
granular materials, even in the absence of particle-shape anisotropy or boundary-driven effects, 
extending beyond previous simulation-based studies. Such geometry-induced convection can 
play a crucial role in mixing, segregation, and particle migration in granular media, with 
direct relevance to natural processes and industrial applications. \\

\noindent
DATA AVAILABILITY \\
The data that support the findings of this study are available from the 
corresponding author upon reasonable request.

\nocite{*}
\bibliography{Refs-Granular}

\end{document}